\documentstyle[aps,prd,twocolumn,epsf,epsfig]{revtex}

\begin{document}


\title{\bf Microscopic Evolution of a Weakly Interacting 
Homogeneous Bose Gas}

\author{D. G. Barci$^{1,2}\;$\thanks{E-mail: barci@physics.uiuc.edu . 
$^{2}\;$Permanent address.}$\;$, E. S. Fraga$^{3}\;$
\thanks{E-mail: fraga@bnl.gov}$\;$,
Rudnei O. Ramos$^{2,4}\;
$\thanks{E-mail: rudnei@peterpan.dartmouth.edu.
$^{2}\;$Permanent address.}}

\address{
{\it $^{1}\;$  Department of Physics, University of Illinois at 
Urbana-Champaign,}\\
{\it 1110W. Green St., Urbana, IL  61801-3080, USA}\\
{\it $^{2}\;$ Universidade do Estado do Rio de Janeiro,
Instituto de F\'{\i}sica,}\\
{\it Departamento de F\'{\i}sica Te\'orica,}\\
{\it 20550-013 Rio de Janeiro, RJ, Brazil} \\
{\it $^{3}\;$ Department of Physics, 
Brookhaven National Laboratory,}\\
{\it  Upton, NY 11973-5000, USA}\\
{\it $^{4}\;$ Department of Physics and Astronomy, Dartmouth College,}\\
{\it  Hanover, New Hampshire 03755-3528, USA}}

\maketitle

\begin{abstract}

We provide a detailed description of the nonequilibrium time 
evolution of an interacting homogeneous Bose-Einstein condensate. 
We use a nonperturbative in-medium quantum field theory approach 
as a microscopic model for the Bose gas. The real-time dynamics of 
the condensate is encoded in a set of self-consistent equations 
which corresponds to an infinite sum of ladder Feynman diagrams. The 
crucial role played by the interaction between fluctuations for 
the instability generation is thoroughly described.

\end{abstract}




\vspace{0.5cm}
\centerline{\it To appear in Laser Physics (2001),}
\centerline{\it Special Issue on Bose-Einstein 
Condensation }
\centerline{\it of Trapped Atoms}

\section{Introduction}

Bose-Einstein condensation (BEC) is a remarkable phenomenon that has been 
present amongst the whole spectrum of research in theoretical 
physics since its proposal\cite{history}. 
It has challenged experimentalists with 
its evasive nature until about five years ago, when it was 
finally observed in a series of elaborate experiments with 
weakly-interacting atomic vapours confined in magnetic traps and 
cooled down to the realm of temperatures of the order of fractions 
of microkelvins \cite{experiments}. 
Such an astonishing environment was made possible 
by the combination of refined techniques such as laser cooling, 
magneto-optical trapping and evaporative cooling. Since this 
remarkable experimental accomplishment, a great amount of 
theoretical investigation has been stimulated (see \cite{trento} 
for a comprehensive review). In part, this interest grew up  because several 
experimental features can be determined by fine-tuning various interesting 
parameters with a high level of control and accuracy. 
Moreover, if one really intends to describe the actual evolution of 
the condensate formation, then finite-density, non-zero temperature, 
and nonequilibrium dynamics effects will have to be taken into 
consideration. This may be accomplished quite naturally 
by in-medium nonequilibrium quantum field theory methods.
This set of characteristics makes BEC one of the most attractive and 
promising systems in which one can use models and approximations 
that could also prove useful in very different environments such as 
neutron stars or heavy-ion collisions \cite{rhic}.

Recent experiments with dilute atomic gases \cite{growth} were able 
to start probing quantities which are relevant to the understanding 
of the underlying dynamics of BEC, such as the time scales for the 
condensate formation and its final size. On the theoretical side, the 
microscopic behavior of such an environment can be appropriately described 
by the nonequilibrium Schwinger-Keldysh closed time-path formalism in 
the quantum field theory approach \cite{keldysh,lebellac}. 
In fact, the first steps in this direction were performed by 
Stoof \cite{stoof}, and provided a qualitative idea of the various 
times scales involved in the BEC process. However, a microscopic 
approach to the condensate onset and growth which can explain the 
mechanism of instability generation and describe the time evolution 
that follows in a quantitative way is still crude and incomplete, even 
in the case of a homogeneous gas. The second stage in the whole process, 
i.e. the growth of the condensate, was considered by Gardiner 
{\it et al.}\cite{gardiner} through a quantum kinetic formalism and 
the construction of a master equation for a density 
operator describing the state of the condensate, which
is equivalent to a Boltzmann equation describing a quasi-equilibrium 
growth of the condensate. Nevertheless, as pointed out by Stoof 
\cite{stoof}, the simplest formulations based on kinetic 
theory do not allow for the observation of a 
macroscopic occupation of the one-particle ground
state, and the question of the instability of the Bose gas system in the 
homogeneous case is a nontrivial one.

In this article, we undertake the task of providing the simplest yet 
sensible nonequilibrium in-medium quantum field theory description 
of the dynamics of condensation of an interacting homogeneous 
Bose-Einstein gas that is abruptly quenched from a temperature far 
above to a temperature far below the critical one \cite{note}. This approach 
yields a nonperturbative set of self-consistent equations which 
correspond to a resummation of the ladder Feynman diagrams discussed 
in ref. \cite{stoof}. In fact, this  sum  
can be interpreted in a simpler, non-diagrammatic way by a careful 
analysis of the role played by the fluctuations, which shows that 
their interaction are crucial to the  
mechanism of instability generation in a model with no instabilities 
at the mean field level. Moreover, as will be described 
below, one can separate the $k$-space into three regions that differ 
in their stability properties, the scale being determined by the 
interaction strength and the density of condensate.
 
The article is organized as follows: Section II presents the 
microscopic model Lagrangian density and an effective theory 
for the fluctuations. Section III contains the self-consistent 
integro-differential equations for the condensate evolution that 
follow from the nonequilibrium approach. Besides, there is a 
detailed discussion of the assumptions and approximations that one 
has to make in order to provide an analytic treatment of the problem. 
In Section IV, we discuss our results for the density of condensate as a 
function of time and present our final comments.

\section{Effective Lagrangian density for the fluctuations}

The interaction Hamiltonian that describes a homogeneous gas of 
interacting bosons 
of mass $m$ has the following second-quantized form (throughout this paper 
we use units such that $\hbar=1$):
\begin{equation}
H=\frac{1}{2}\int d^3 x~ d^3 x' 
\phi^\dagger (\vec{x'},t)\phi^\dagger (\vec{x},t)
V(\vec{x}-\vec{x'})\phi(\vec{x'},t)\phi(\vec{x},t) \, ,
\label{hamiltonian}
\end{equation} 
where $\phi(\vec{x},t)$ and $\phi^{\dagger}(\vec{x},t)$ are the boson  
anihilation and creation Heisenberg field 
operators, and $V(\vec{x}-\vec{x'})$ is the two-body interatomic potential. 
Since we will consider a weakly-interacting dilute gas, we 
are allowed to use the fact that, in this case, only binary collisions 
at low energy are relevant, so that we can approximate the actual 
interatomic potential by the much simpler form
\begin{equation}
V(\vec{x}-\vec{x'})=g \delta(\vec{x}-\vec{x'}) \quad ,
\label{potential}
\end{equation}
where the coupling constant $g$ is related to the $s$-wave 
scattering lengh $a$ through
\begin{equation}
g = \frac{4\pi a}{m} \quad .
\label{lambda}
\end{equation}
Then, for a weakly-interacting dilute and cold gas, 
we end up with a hard core interaction potential. In fact, one is 
reasonably safe to use such approximation since typical values 
for the dimensionless parameter which controls the validity of the 
dilute-gas approximation, the number of particles in a 
scattering volume, i.e.  $n |a|^3$, is always less than $\sim 10^{-2}$. 
Inspired by the discussion above, one can  
write a field theory model for this system. It corresponds to the 
simplest model for a nonrelativistic complex Bose field, and its  
Lagrangian density is given by
\begin{equation}
{\cal L}=\phi^*\left( i\frac{d}{dt}+\frac{1}{2m}\nabla^2\right)\phi
+\mu \phi^* \phi -\frac{g}{2}(\phi^*\phi)^2 \quad ,
\label{lagrangian}
\end{equation}
where the complex scalar field $\phi({\bf x},t)$ represents charged spinless 
bosons of mass $m$, and $g$ is the coupling constant defined above. 
In (\ref{lagrangian}) we have also explicitly introduced a 
chemical potential $\mu$ that guarantees a constant total density of 
particles 
\begin{equation}
\langle\phi^*\phi\rangle =n \quad .
\label{density} 
\end{equation}
We also assume that the system 
is coupled to a heat bath environment with inverse temperature 
$\beta=1/(k_B T)$. The role played by temperature is very vague at this 
point. It will be discussed in detail later. For the moment, it is here 
just to remind us that we will consider the nonequilibrium evolution of 
the system at finite density and {\it nonzero} temperature.

Since we are interested in describing the onset and the time evolution 
of the Bose-Einstein condensate, it is convenient to decompose the 
original fields $\phi$ and $\phi^*$ into a condensate (zero-momentum mode) 
part given by $\varphi_0$ and $\varphi_0^*$, and a fluctuation part, 
$\varphi$ and $\varphi^*$, that accounts for the atoms outside the 
condensate. Assuming a homogeneous condensate, we define the following 
decomposition:
\begin{eqnarray}
\phi(\vec{x},t)&=&\varphi_0(t)+\varphi(\vec{x},t) \quad ,\nonumber \\
\phi^*(\vec{x},t)&=&\varphi_0^*(t)+\varphi^*(\vec{x},t) \quad .
\label{shift}
\end{eqnarray}
Note that we make a time-dependent shift, i.e., we take $\varphi_0(t)$ 
as an arbitrary function of time that will be determined by the 
dynamics of the system. This procedure is different from the usual 
equilibrium analysis of systems that feature spontaneous symmetry 
breaking, where one usually performs a constant shift. Since we intend 
to describe the nonequilibrium evolution of the condensate, we must 
keep the time dependence.

The substitution of the shifted fields (\ref{shift}) in the Lagrangian 
(\ref{lagrangian}) upto second order yields the Bogoliubov approximation 
for quasiparticles, which neglects the interaction between fluctuations 
and is reasonable only at temperatures well below the critical temperature 
for the condensate formation. At this level of approximation, the system 
is absolutely stable, and this corresponds to the stationary equilibrium 
situation that is achieved after the whole dynamical evolution of the 
condensate has taken place. In order to allow for the appearance of an 
instability, which will be responsible for the onset and subsequent 
evolution of the condensate as a function of time, we must go beyond the 
Bogoliubov approximation. The simplest extension is to implement a mean 
field approximation in the interactions between the fluctuation fields. 
It turns out that even this simple improvement in our approach is able 
to trigger the instability that will drive the condensate formation. The 
mean field approximation corresponds to a self-consistent Hartree approach, 
and amounts to the following modification in the interaction term for 
fluctuations:
\begin{eqnarray}
g (\varphi^* \varphi)^2&=&
4 g \langle \varphi^* \varphi \rangle
\varphi^* \varphi + 
g \langle \varphi^* \varphi^* \rangle
\varphi \varphi +
g \langle \varphi \varphi \rangle
\varphi^* \varphi^* +  
\nonumber \\
&+& \left[ g(\varphi^* \varphi)^2 -
4 g \langle \varphi^* \varphi \rangle
\varphi^* \varphi - \right. \nonumber \\
&-& \left. g \langle \varphi^* \varphi^* \rangle
\varphi \varphi -
g \langle \varphi \varphi \rangle
\varphi^* \varphi^*   
\right] \quad ,
\label{meanfield}
\end{eqnarray}
where the first terms in the rhs is taken as part of the 
quadratic Lagrangian for fluctuations, and the terms inside
the square brackets is taken as part of the interaction 
Lagrangian. The terms $\langle \varphi \varphi \rangle$ and 
$\langle \varphi^* \varphi^* \rangle$ are usually known as the
anomalous density terms while $\langle \varphi^* \varphi \rangle$
is the non-condensate density. In the following
we work with the approximation of neglecting the anomalous
density terms in (\ref{meanfield}), which is 
equivalent to the the Hartree-Fock-Bogoliubov-Popov (HFBP)
(or dynamical Popov) approximation for dilute Bose gases, which is
generically valid for $T \ll T_c$, when the contributions
from the anomalous and non-condensate densities are very small as
compared to the condensate density. This approach 
has also shown to give a reasonable first approximation for the description of
the excitation spectrum at all temperatures provided the non-condensate density
is calculated self-consistently \cite{griffin2}. 

{}From the decomposition of the fields and
(\ref{meanfield}) in the HFBP approximation, 
the quadratic part of the Lagrangian density for
the fluctuations, ${\cal L}_0 (\varphi,\varphi^*)$, may then be written as
\begin{eqnarray}
{\cal L}_0 (\varphi,\varphi^*) &=& \varphi^*\left[ i\frac{d}{dt}+\frac{1}{2m}
\nabla^2 \right]\varphi + \nonumber \\
&+&\varphi^* \left(-\frac{g}{2} \varphi_0^2 \right) \varphi^* 
+ \varphi \left(-\frac{g}{2} {\varphi_0^*}^2 \right) \varphi \quad .
\label{eq2}
\end{eqnarray}
Here, we have used the fact that, under the field decomposition in 
the condensate and out of the condensate modes, the density constraint 
then becomes 
\begin{equation}
\langle \phi^* \phi \rangle =
|\varphi_0|^2 + \langle \varphi^* \varphi \rangle = n \quad .
\label{constraint}
\end{equation}
Additionally, assuming that at the initial time the system is mostly composed
of particles outside the condensate, $\langle \phi^* \phi \rangle \simeq
\langle \varphi^* \varphi \rangle$ (at $t=0$), simple relations
involving the generating functional for the correlation functions
(see, for instance, the last section of chapter 2 in \cite{baym})
allow us to write the total number density 
$n$ of particles in terms of the chemical potential $\mu$, valid in the
mean-field approximation for the potential, as:
$\mu = 2 g n$. This expression does not depend on any equilibrium property 
of the system, depending only on the mean-field approximation. 
Notice that this value for $\mu$ coincides with the usual one  
obtained in the equilibrium  treatments of dilute Bose gases in the 
HFBP approximation \cite{griffin2}. Moreover, it satisfies the 
Hugenholtz-Pines relation \cite{pines} that would be obtained in the 
equilibrium problem. In spite of this, our derivation is valid for the 
out-of-equilibrium regime and should not to be confused with the 
equilibrium treatments. These considerations lead to the quadratic
Lagrangian for the fluctuations shown above (Eq. (\ref{eq2})). 

\section{The quench scenario, Green functions and self-consistent equations}

The scenario proposed here assumes that, for $t<0$, the initial state 
is in {\it equilibrium} at a temperature $T_i \gg T_c$. At
$t=0$ the system is then abruptly quenched
to a much lower temperature
$T_f \ll T_c$, where $T_f$ is the temperature of the thermal bath
in which the system is immersed (the one associated with the 
quantity $\beta$ that appears in our description) and, 
of course, it will be the
equilibrium temperature which the system will reach asymptotically 
as $t\to\infty$. 
We should remark, then, that the initial temperature itself is not 
important in our study and it will not appear in our equations. It is 
only a conceptual tool that will help us in the definition of the 
initial conditions for the number densities, as will be clear below. 
In fact, since our approach is strongly out of equilibrium, it is not 
possible to define a ``temperature parameter'' when the condensate is 
evolving as it is usually done in a quasi-equilibrium treatment. 
Technically, this means that the final {\it equilibrium} 
temperature, $T_f=1/(k_B \beta)$, will 
show up in the nonequilibrium propagators through boundary 
conditions (see below).
 
This kind of quench is easily attained in the experiments
of Bose-Einstein condensation of atomic gases, where the typical
relaxation time scales are long enough (around $\sim 0.1 s$,
depending on the temperature \cite{MIT}) to allow for a fast drop
in the temperature of the system that evolves afterwards
out of equilibrium. With
this choice of initial state, it is reasonable to consider the
dynamics of the building up of the condensate state, which at
the initial time is $n_{\rm cond.}(t=0) = |\varphi_0(t=0)|^2 \approx
0$,
and the depletion of the excited states, which at $t=0$ is
given by $n_{\rm exc.}(t=0) =\langle
\varphi^* \varphi \rangle \approx n$, as essentially a two-level 
problem. It is clear that this approximation breaks down
for temperatures close to the critical temperature, where
the detailed treatment would require a thorough study
of the dynamics among the many levels of excited states. 
Furthermore, at 
this point the mean field approximation would not be reliable any more.

In the scenario described above, the condensate builds up
subject to the density constraint relation, which may be
expressed in terms of the averages of the real and the imaginary
parts of $\varphi$ and $\varphi^*$
($\varphi = \varphi_1 + i \varphi_2$ and $\varphi^* 
= \varphi_1 - i \varphi_2$). Spatial translation 
invariance yields:
\begin{eqnarray}
\lefteqn{|\varphi_0 (t)|^2 + n_{\rm exc.}(t) = n \;,} \nonumber \\
& & n_{\rm exc.}(t) = \langle
\varphi_1 (t) \varphi_1 (t) \rangle_{\varphi_0} + \langle
\varphi_2 (t) \varphi_2 (t) \rangle_{\varphi_0} \quad .
\label{dens2}
\end{eqnarray}

In order to express these self-consistent equations in a concrete and 
convenient form let us  define the field averages in terms of the Green 
functions
for the fields. The Green functions can be defined from a generating
functional for our Lagrangian model, Eq. (\ref{lagrangian}), 
where the generating functional $Z[J]$, in terms of an external source $J$,
is given by

\begin{equation}
Z[J] = \int_C D \phi D\phi^* \exp \left\{ i S[\phi,\phi^*,J]\right\} \:,
\label{ZJ}
\end{equation}

\noindent
where the classical action is given by

\begin{equation}
S[\phi,\phi^*,J] = \int_C d^4 x \left\{ {\cal L}[\phi,\phi^*] + J(x) \phi^*(x)
+ J^*(x) \phi(x) \right\} \: .
\label{SphiJ}
\end{equation}

\noindent
In (\ref{SphiJ}) the time integration is along a contour suitable for 
real-time evaluations, which we choose as being the Schwinger closed-time
path \cite{keldysh,lebellac}, of common use in condensed matter 
non-equilibrium problems, where the time path $C$
goes from $-\infty$ to $+\infty$ and then back to $-\infty$. The functional
integration in (\ref{ZJ}) is over fields along this time contour. As in
the Euclidean time formulation, the scalar field is still periodic in time,
but now $\phi(t, \vec x)=\phi(t-i\beta, \vec x)$. Temperature appears
due to the boundary conditions, but now real time is explicitly present in the
integration contour.
Denoting by $J^+$ and $J^-$ the sources in the $-\infty$ to $+\infty$
path and $+\infty$ to $-\infty$ paths, respectively, along the time path
contour, the generating functional
$Z[J^+, J^-]$ can be written as

\begin{eqnarray}
Z[J^+,J^-] & = & \int D \phi D \phi^*
\exp\left(i\int d^4 x\left\{{\cal{L}}[\phi_+,\phi_+^*,J^+]
\right. \right. \nonumber \\
&-& \left. \left.
{\cal{L}}[\phi_-,\phi_-^*,J^-]\right\}\right.\Bigr)\;.
\label{functional}
\end{eqnarray}
By performing the path integrals over the quadratic forms as usual, 
one obtains that the generating functional in the
Schwinger's closed-time path formalism can be written as
(with $\phi$, $\phi^*$ written in terms of the real fields
$\phi_1$ and $\phi_2$)

\begin{eqnarray}
Z[J^+,J^-] \!\! &=& \! \!\exp\left\{i\int d^4 x \!\! \left[
{\cal{L}}_{int}\left(\!-i\frac{\delta}{\delta J^+_j}\!\right) \!-\!
{\cal{L}}_{int}\left(i\frac{\delta}{\delta
J^-_j}\!\right)\right] \right\}  \nonumber \\
           &   & \times \exp\left\{\!\frac{i}{2}\! \int d^4 x ~d^4 y~
\!\! J^a_j(x) G^{ab}_{jj} (x,y) J^b_j (y)
\right\}
\label{function1}
\end{eqnarray}
with $a,b = +,-$ and $j=1,2$.
The Green functions that enter the integrals along
the closed-time path contour in (\ref{function1}) are given by 
\cite{lebellac,stoof} 
(in momentum space)
\begin{eqnarray}
&G_{jj}^{++}& ({\bf k},t,t')  =  G_{jj}^> ({\bf k},t,t') \theta(t-t')+
G_{jj}^< ({\bf k},t,t') \theta(t'-t)\; , \nonumber \\
&G_{jj}^{--}& ({\bf k},t,t')  =  G_{jj}^> ({\bf k},t,t') \theta(t'-t)+
G_{jj}^< ({\bf k},t,t') \theta(t-t')\; , \nonumber \\
&G_{jj}^{+-}& ({\bf k},t,t')  =  -G_{jj}^< ({\bf k},t,t')\; , \nonumber \\
&G_{jj}^{-+}& ({\bf k},t,t')  =  -G_{jj}^> ({\bf k},t,t') \quad .
\label{greens}
\end{eqnarray}
The functions  $G^>$ and $G^<$ satisfy the property
$G^<_{jj} ({\bf k},t,t') = 
G^>_{jj} ({\bf k},t-i \beta,t')$, which is recognized as the
periodicity condition in real time, Kubo-Martin-Schwinger (KMS)
condition. Here, $\beta$ is the inverse of the temperature of the
thermal bath and appears as a consequence of the  
boundary conditions arising from the construction of the complex time 
path.

The field averages of last section can then be expressed in terms of the
Green functions for $\varphi_1$ and $\varphi_2$ as ($j=1,2$)
\begin{equation}
\langle \varphi_j (t) \varphi_j (t) \rangle = \int \frac{d^3 k}{(2 \pi)^3}
\left[-i G_{jj}^> ({\bf k},t,t) \right] \quad ,
\label{average}
\end{equation}

$G^>$ and $G^<$ are constructed from the homogeneous solutions
to the operator of quadratic fluctuation modes which, using Eq. (\ref{eq2})
expressed in terms of $\varphi_1$ and $\varphi_2$, are given by
(in momentum space)
\begin{eqnarray}
\lefteqn{\frac{d \chi_2 ({\bf k},t)}{d t} + \left(\frac{{\bf k}^2}{2 m} + 
g |\varphi_0|^2 \right) \chi_1 ({\bf k},t)= 0}  \nonumber \\
& & \frac{d \chi_1({\bf k},t)}{d t} - \left(\frac{{\bf k}^2}{2 m} - 
g |\varphi_0|^2 \right) \chi_2 ({\bf k},t)= 0 \quad .
\label{modes}
\end{eqnarray}

The boundary conditions for the solutions of the equations 
above are such that, for $t<0$, $|\varphi_0 (t)|^2=0$ and 
$\chi_1 ({\bf k},t)$ and $\chi_2 ({\bf k},t)$ can, for example, be given by
$\chi_1 ({\bf k},t) = e^{i\varepsilon_{\bf k}t}/\sqrt{2 N}$ and 
$ \chi_2 ({\bf k},t)=i e^{i\varepsilon_{\bf k}t}/\sqrt{2 N}$, where 
$\varepsilon_{\bf k}={\bf k}^2/(2 m)$
and the normalization factor $N$
can be fixed by imposing the constraint condition, 
Eq. (\ref{constraint}), at $t=0$ ( $|\varphi_0 (0)|^2=0$).
We should stress that, from our choice of initial configuration, we have  
a well-defined initial state, given by $n_0 \simeq 0$ 
and $n_{exc} \simeq n$, and a well-defined equilibrium
final state, at temperature $T_f \ll T_c$, towards which the system 
will evolve. The final situation, then, is an equilibrium one, which  
allows for the definition of a final equilibrium density matrix, 
$\rho = Tr ~e^{-\beta H}$, where $\beta$ is the inverse of the 
temperature of the heat bath that was responsible for the quench 
($T_f \ll T_c$). Therefore, one can 
consistently define the Schwinger-Keldysh closed time-path procedure in 
order to describe the nonequilibrium evolution that will follow. 
The conditions 
fulfilled by the nonequilibrium propagators in that formalism are 
analogous to the KMS conditions obtained in the imaginary time 
description of equilibrium problems.

By decoupling the set of equations in (\ref{modes}), we obtain

\begin{eqnarray}
&&\ddot{\chi}_1  - \frac{\dot{\omega}_-}{\omega_-} \dot{\chi}_1 
+ \omega_+ \omega_- \chi_1 =0 \nonumber \\
&&\ddot{\chi}_2  - \frac{\dot{\omega}_+}{\omega_+} \dot{\chi}_2
+ \omega_+ \omega_- \chi_2 =0 \;,
\label{modes2}
\end{eqnarray}

\noindent
where $\omega_\pm = {\bf k}^2/(2m) \pm g | \varphi_0|^2$. 
Defining $\xi_1 = \chi_1/\sqrt{\omega_-}$ and $\xi_2 = \chi_2/\sqrt{\omega_+}$,
Eq. (\ref{modes2}) can also be written as

\begin{eqnarray}
&&\ddot{\xi}_1 + \omega_1^2 \xi_1 =0 \nonumber \\
&&\ddot{\xi}_2 + \omega_2^2 \xi_2 =0 \;,
\end{eqnarray}

\noindent
where

\begin{equation}
\omega_{1(2)}^2 = \frac{1}{2}  \frac{\ddot{\omega}_{-(+)}}{\omega_{-(+)}}
- \frac{3}{4} \frac{\dot{\omega}_{-(+)}^2}{\omega_{-(+)}^2}+
\omega_+ \omega_- \;.
\label{omega12}
\end{equation}

\noindent
We can see from these decoupled equations  that
those modes with ${\bf k}^2/(2m) < g
|\varphi_0|^2$ are the  unstable ones and they will drive initially the 
excited particles
towards condensation. At this point, 
one should recall that the instability arises as a 
consequence of the interaction between fluctuations. Should one 
discard those interactions, the mean-field description would result 
in a absolutely stable system. This situation is completely 
different from the one encountered in the relativistic approach \cite{boy}.
Those unstable modes, 
for which $k^2/(2m)$ is smaller than  $g |\varphi_0|^2$,
correspond to the (exponential) growth of the long-wavelength 
fluctuations which drive the process of phase transition,
or condensation in our case. This is similar to the phenomenon of
spinodal decomposition in statistical mechanics, typical of
second order phase transitions \cite{spinodal}.

In terms of the mode functions $\chi_1$ and $\chi_2$, and taking into 
account the boundary conditions for them, the Green's 
functions can be expressed as ($j=1,2$)

\begin{eqnarray}
G_{jj}^> ({\bf k},t,t') &=& \frac{i}{1-e^{-\beta \epsilon_{\bf k}}}
\left[ \chi_j ({\bf k},t)\chi_j^* ({\bf k},t') + \right.\nonumber \\
&+& \left. e^{-\beta \epsilon_{\bf k}}
\chi_j^* ({\bf k},t)\chi_j ({\bf k},t')\right]\;.
\label{Gmodes}
\end{eqnarray}

\noindent
The Green's functions for the complex
fields $\varphi,\varphi^*$ are expressed as usual by \cite{stoof}:
\begin{equation}
\langle \varphi (t) \varphi^* (t) \rangle = \int \frac{d^3 k}{(2 \pi)^3}
\left[-i G_{\varphi \varphi^*}^> ({\bf k},t,t) \right] \;,
\label{averagec1}
\end{equation}
\begin{equation}
\langle \varphi^* (t) \varphi (t) \rangle = \int \frac{d^3 k}{(2 \pi)^3}
\left[-i G_{\varphi^* \varphi}^> ({\bf k},t,t) \right] \;.
\label{averagec2}
\end{equation}

In terms of (\ref{averagec1}) and (\ref{averagec2}), we have that the 
non-condensate density $n_{\rm exc}(t)$ can be written as
\begin{eqnarray} 
n_{\rm exc} (t) &=& \langle\varphi_1 (t) \varphi_1 (t) \rangle + \langle
\varphi_2 (t) \varphi_2 (t) \rangle = \nonumber \\
&=& \frac{1}{2}(\langle
\varphi (t) \varphi^* (t) \rangle + \langle
\varphi^* (t) \varphi (t) \rangle) \quad ,
\label{2field}
\end{eqnarray}
and the KMS condition can be 
expressed, in this case, as $G_{\varphi \varphi^*}^> ({\bf k},t-i \beta,t')
=[G_{\varphi^* \varphi}^> ({\bf k},t,t')]^*$, or 
$G_{\varphi^* \varphi}^> ({\bf k},t-i \beta,t')
=[G_{\varphi \varphi^*}^> ({\bf k},t,t')]^*$.

Using Eq. (\ref{Gmodes}),
we then obtain that $ n_{\rm exc}(t)$ can be expressed as 
\begin{eqnarray}
n_{\rm exc}(t) &=& 
\int \frac{d^3 k}{(2 \pi)^3} \left[|\chi_1({\bf k},t)|^2 +\right.\nonumber \\
&+& \left. |\chi_2 ({\bf k},t)|^2 \right] \coth\left(\frac{\beta \epsilon_{\bf k}}{2} \right) \quad .
\label{integral}
\end{eqnarray}

At $t=0$, from the initial conditions imposed over the system, the system
is all in the non-condensate state and the condensate density
$n_c \equiv |\varphi_0|^2=0$; $\chi_1$ and $\chi_2$ are purely oscillatory
functions. We then obtain that 

\begin{equation}
n_{\rm exc}(0) \equiv n= \langle \varphi^* \varphi \rangle\bigr|_{t=0} \equiv
\langle \varphi_1^2 \rangle\bigr|_{t=0} + \langle \varphi_2^2 \rangle\bigr|_{t=0}
\;,
\label{constraint2} 
\end{equation}
which, from Eq. (\ref{integral}), gives

\begin{eqnarray}
n= \frac{1}{2 N \pi^2}
\int dk~ k^2 \coth\left(\frac{\beta \varepsilon_{\bf k}}
{2} \right) \quad .
\label{integral2}
\end{eqnarray}

\noindent
The $T=0$ part of (\ref{integral2}) is divergent and represents the 
zero-point energy. By subtracting the zero-point energy we obtain  
for the normalization $N$
\begin{equation}
N = \frac{2}{n}\zeta(3/2) \left(\frac{m}{2 \pi
\beta}\right)^{\frac{3}{2}}\;.
\label{norm}
\end{equation}
Using the expression for the density in terms of the critical temperature
for condensation of an ideal Bose gas ($1/\beta_0$) \cite{griffin}, $n=
\zeta(3/2) \left[m/(2 \pi
\beta_0)\right]^{3/2}$, we find a
simple expression for the normalization $N$ as given by $N= 2
(\beta_0/\beta)^{3/2}$, which corresponds to the pre-factor 
used in Ref. \cite{BFR}.
Using these constraint conditions together with the initial boundary 
conditions, from Eq. (\ref{integral}), we can then express the condensate 
density as

\begin{eqnarray}
|\varphi_0(t)|^2 &=& \frac{1}{2\pi^2} \left(\frac{\beta}{\beta_c}\right)^{3/2}
\int_0^{\sqrt{8\pi a|\varphi_0(t)|^2}}
dk~k^2~ \times \nonumber \\
&\times& \left[ 1-
N \left(|\chi_1({\bf k},t)|^2 + |\chi_2({\bf k},t)|^2\right)\right]
n_{{\bf k}}(\beta) \quad ,
\label{vinculo}
\end{eqnarray}

\noindent
where $n_{{\bf k}}(\beta)=(e^{\beta\epsilon_{{\bf k}}}-1)^{-1}$. 
The ultraviolet
divergences were explicitly cancelled by subtracting 
the 
$T=0$ component of (\ref{integral}). This is fine, since the 
contributions from the non-condensate part of the density 
are neglegible for weak interacting  dilute Bose gas systems 
at $T=0$ \cite{trento}. 
Note also that, in our out-of-equilibrium approach, 
there are no infrared divergences since the finite time is 
a natural regulator. However, for the equilibrium $t\rightarrow\infty$,
the critical temperature 
will be modified by the interactions as pointed out in Ref.
\cite{GB,CPR}. 

Equations (\ref{modes}) and (\ref{vinculo}) form an 
integro-differential system that may be solved 
for $\varphi_0(t)$ numerically, given the initial 
conditions for $\varphi_0(t)$, 
$\chi_1({\bf k},t)$ and $\chi_2({\bf k},t)$ mentioned before. Indeed, this 
system of equations determines completely the time evolution of the 
condensate density as a function of the temperature and of the total density 
of the gas.
At this point, one could ask about the contribution from those terms
responsible for the decay of quasiparticles which are
related, microscopically, to the imaginary part of
the self-energies in higher order perturbation theory.
These contributions are important to describe the detailed
dynamics for temperatures close to the critical temperature, 
which is not the case we have at hand. In such range of temperatures, 
the typical time scales for condensation are long
enough so that scattering and growth among adjacent levels,
which are described by those microscopic processes, are of
relevance to the dynamics in this regime. Those
processes have been described in the literature through
the use of quantum kinetic theory, i.e. quantum Boltzmann equations, 
and could as well be done within the approach described here  
when considering higher-order corrections to our field
averages.

Equation (\ref{integral}) is the first order term in the finite-temperature
quantum field perturbation expansion for $\langle \varphi_j^2 \rangle$.
Higher-order corrections for the equal-time two-point field
averages can be  expressed
in terms of the coincidence limit of the (causal)
two-point Green functions $G_{\varphi \varphi^*}$ and 
$G_{\varphi^* \varphi}$, which satisfy
the Dyson equation (the indices stand for $\varphi$ and $\varphi^*$):
\begin{equation}
G_{ij}=G_{ij}^0 + G_{ik}^0\Sigma_{kl} G_{lj} \quad ,
\label{fullG}
\end{equation}
where $\Sigma_{ij}$ is the (matrix) self-energy, and $G_{ij}^0$ is the
zeroth-order non-interacting Green function,
satisfying the equation (in momentum space) 
\begin{equation}
\left[ \pm i \frac{d}{dt} - \varepsilon_{\bf k} \right] 
G^0_{\varphi \varphi^*(\varphi^* \varphi)}({\bf k},t,t') = \delta(t-t) \quad .
\label{G0}
\end{equation}
One of the advantages of expressing the Green's functions in terms
of the solutions of (\ref{modes}) is the possibility of obtaining, 
in an unambiguous way, all higher-order corrections to the two-point
and many-point functions.


\section{Results and discussion}

{}From Eq. (\ref{omega12}) we find that initially the 
dynamics will be dominated by essentially  
two regimes in the $k$-space: for $(k^2/2m) \ll g|\varphi_0|^2$, or
spinodal regime, is the
regime caracterized by the exponential growth of the fluctuations, 
driving the system towards Bose-Einstein condensation.
{}Following a crossover at $(k^2/2m) \sim g|\varphi_0|^2$, the modes for 
which $(k^2/2m) \gg g|\varphi_0|^2$ will be oscillatory stable modes.
The energy of these modes are high enough such that they effectively
sample a symmetric potential and will not contribute to the dynamics.
These properties determined our choice of momentum cut-off in 
Eq. (\ref{vinculo}). 
Equations (\ref{modes}) and 
(\ref{vinculo}) form a set of integro-differential equations
that can be solved numerically, given the initial conditions
mentioned before. In {}Fig. 1 we show the results for 
the condensate density $|\varphi_0|^2$, as a function of time,
for different values of temperature. {}From these results we notice
the characteristic growth curves for the condensate. {}For very initial times, 
the system is dominated by the stable modes. As the condensate develops 
the number of unstables modes increases. At some point they dominate and
make the condensate grow exponentially fast, until the equilibrium 
state is reached. This state is determined by the temperature at which the 
system is quenched below the critical temperature, and by the total density.

It is important to point out that the evolution of the 
condensate is completely driven by the interaction between the microscopic 
fluctuations of the field around the condensate. In fact, in this 
non-relativistic system, the interaction 
between fluctuations is responsible for the appearance of the instability 
which allows for the onset of the condensate (see, {\it e.g.}, \cite{boy} 
for a description of what happens in the relativistic case). 
One should also note that not all the excited
particles condense, since there will always be a fraction
(which depends on various parameters for a particular
system, and on the temperature of the thermal bath)
of excited modes, with high enough frequency, that 
remains stable.

\begin{figure}[htp]
\centerline{\hbox{\epsfig{figure=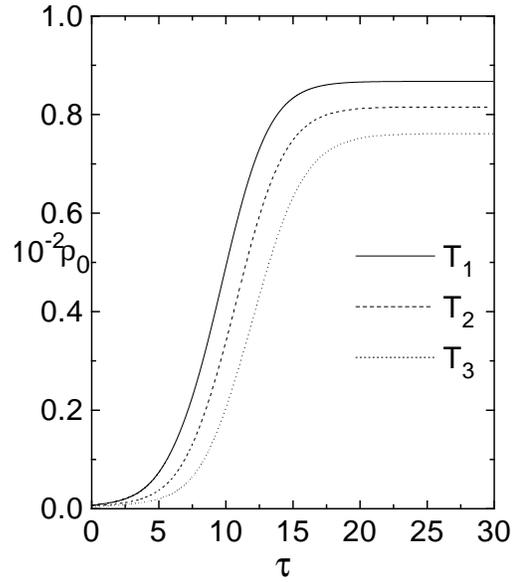,width=7cm}}}
\caption{Condensate density as a function of time for $na^3=0.01$ and
$T_1/T_0=0.06$, $T_2/T_0=0.08$ and $T_3/T_0=0.1$. Here, 
$\tau\equiv (\hbar/ma^2)t$ is a dimensionless time and 
$\rho_0 \equiv a^3 |\varphi_0|^2$ is a dimensionless density.}
\end{figure}

The system of equations we have derived is completely nonperturbative
and able to describe the time evolution of BEC in the extremelly far from 
equilibrium regime, where standard methods based on the mean-field
Gross-Pitaevski equations are not applicable. In this sense, our description
should complement the usual approach via Boltzmann 
equation.
{}Furthermore, the system of equations obtained here, can be viewed 
as a resummation of the ladder Feynman diagrams mentioned by 
Stoof in \cite{stoof}.

In summary, we have developed an out-of-equilibrium 
non-perturbative quantum field theory description of the condensation 
process of an interacting homogeneous Bose-Einstein gas quenched below 
the critical temperature. 
Although we focused this article on the instability process that generates 
the condensate ({\it i.e.}, the short time behavior), for $t\rightarrow\infty$ 
our 
results confirm the expected behavior for this limit\cite{stoof}. 
However, the equilibrium 
($t\rightarrow\infty$) values of the condensate fraction are lower 
than the experimental results \cite{MIT} and the calculations of 
Dalfovo {\it et. al.} \cite{trento}. 
This may be due to our approximation of neglecting
incoherent
collisional processes, which is a valid approximation in an infinite 
homogeneous gas at very low temperatures and densities, 
but otherwise may give an important contribution. We expect that the
self-consistent inclusion of pair terms should account for most
of these contributions.

In spite of the absence of non-homogeneity effects, we hope 
that the approach developed here may be useful in the analysis of 
transients in realistic Bose-Einstein condensation experiments with atomic 
gases. Our results should also be applicable to trapped atomic
gases in the central region of wide traps.  
Moreover, with a suitable generalization of the formalism 
presented above, 
we could be able to develop a theoretical description of the 
dynamical aspects of a recently proposed experiment \cite{chiao}, 
regarding the Bose-Einstein condensation in a 
weakly-interacting photon gas in a nonlinear Fabry-Perot cavity 
\cite{alessandro}.


\acknowledgements

The authors acknowledge support from CNPq (D.G.B., E.S.F. and R.O.R.) 
and SR2-UERJ (D.G.B. and R.O.R.).
D.G.B. is partially supported by NSF through grant DMR-9817941.
E.S.F. is partially supported by the U. S. Department of Energy under 
Contract No. DE-AC02-98CH10886. 



\begin{references}
\bibitem{note} Most of the results discussed here were presented in 
a much shorter version in ref. \cite{BFR}.
\end{references}
\end{document}